\providecommand{\keywords}[1]{
  \small\textbf{\textit{Keywords---}} #1
}
\documentclass[a4paper,fleqn,3p]{elsarticle}

\usepackage{hyphenat}
\usepackage{graphicx}

\begin{document}

\title{Characteristics of a matrix proportional counter with circular anodes}

\author[col]{R.A.~Etezov}
\author[col]{Yu.M.~Gavrilyuk}
\author[col,nd]{A.M.~Gangapshev}
\author[col]{V.V.~Kazalov}
\author[col,nd]{A.Kh.~Khokonov}
\author[col,nd]{V.V.~Kuzminov}
\author[hn]{S.I.~Panasenko}
\author[hn]{S.S~Ratkevich
\corref{cor1}}
\ead{ssratkevich@karazin.ua}
\cortext[cor1]{Corresponding author}

\address[col]{Baksan Neutrino Observatory INR RAS, KBR, Russia}
\address[nd]{H.M.~Berbekov Kabardino-Balkarian State University, Russia}
\address[hn]{V.N.~Karazin Kharkiv National University, Ukraine}

\begin{abstract}
The construction of a butt-end multicell matrix proportional counter (MMPC) is presented in the work. Each cell is a butt-end proportional counter with a 5 mm cathode diameter and 0.8 mm circular anode diameter.
In the example of the $3\times3$ matrix, it is shown that the gas multiplication of the central cell depends on the potentials at the anodes of the peripheral cells, the drift electrode, the forming rings, and the surrounding metal parts of the structure.
The amplitude characteristics were measured when the MMPC was filled with mixtures of 96.3\% Ar + 3.7\% Xe and 90\% Ar + 10\% CH$_4$ at pressures of 620 Torr and 62 Torr. The calibration was carried out with $\alpha$-particles and $\gamma$-quanta from a $^{238}$Pu source. For photons with energies of 7.5 keV, a resolution of 26\% was obtained. It is shown that, based on MMPC, it is possible to fabricate recording surfaces of arbitrary configuration.
\end{abstract}

\maketitle
\noindent

\noindent

\section{Introduction}
\label{sec:introduction}
Results of investigations of the two-dimensional Micro-Pixel Chamber
($\mu$-PIC) made on double-sided printed-circuit-board (PCB) technology and
intended for registration of ionization signals in the gas drift chambers ware
published in a set of articles in the last years, for example \cite{Miuchi03,Nagayosh04,Takada07}.
The $\mu$-PIC structure consists of a set of parallel copper cathode strips
located with a pitch of 400 $\mu$m on a polyimide substrate on the side of the
drift volume.
Holes with a diameter of 50 $\mu$m are made in the centre of each cavity in
the substrate.
Copper free cavities with 400 $\mu$m pitch are made along axis of each strip.
Holes on the backside of the substrate are configured along the axis of the
400 $\mu$m pitch anode strips, which is orthogonal to the cathode ones. Holes
are filled with electrodeposited copper. Each core butt end in the center of
the cathode strip cavities is an anode of a proportional counter. Such
structure allows one to reach a good enough spatial resolution ($\sim$200
$\mu$m) and demonstrates a relatively low level of random discharges signals
when working at the gas-multiplication factor 5000.

Investigations have shown that $\mu$-TPC could be used as part of drift
time-projection chambers intended to register three-dimensional images of
tracks from Compton electrons, $\beta$- and $\alpha$-particles, and recoil
nuclei from neutron or hypothetical dark matter particles scattering.
In the latter case, the energy resolution becomes a critical parameter, which
affects the accuracy of reconstructing the energy spectrum of recoil nuclei.
In \cite{Nagayosh04}, for a recording plane with dimensions of $10\times10$ cm$^2$, a
resolution of $\sim30$\% was obtained at an energy of 5.9 keV. In \cite{Ochi2001}, a
resolution of 20\% was obtained for a recording plane with dimensions of
$3\times3$ cm$^2$.

It was assumed that the deterioration in resolution for the larger plane was
due to the increased variability in the shape of anode electrodes due to the
imperfect technology of circuit boards.  However, as the practice has shown,
an energy resolution does not improve with the technology improvement.

One can be assumed that there is another underestimated factor that can lead
to a deterioration in resolution with an increase in the size of the board.
This factor may be the collective influence of the fields of neighbouring
anodes on the value of the gas gain of a particular anode. It is quite
difficult to investigate this issue within the framework of the printed
circuit board configuration due to the rigid mechanical connection of the
elements. It seems optimal to carry out such a study on a matrix plane
consisting of separate, independent cells.

Investigation of the flowing open butt-end proportional counter working
characteristics was done recently in the Baksan Neutrino Observatory of the
INR RAS. A circular butt-end of a medical needle with 0.8 mm diameter was used
as an anode. A blowing off was implemented through the inner needle channel.
The counter was intended for a registration of low energy electrons from the
low-active tagged biological samples. The energy resolution of the 5.9 keV
X-ray was found to be $\sim$34\% for the 95\% Ar + 5\% Xe gas mixture at 620
Torr pressure.

The counter showed good operating stability characteristics for a set of the
gas mixtures at the 62-620 Torr pressures. The working characteristics
reconditioned without consequences after a high voltage decreasing at a time
of an amplitude characteristics measurement if a discharge voltage was
reached.
The analysis of the potential possibilities, manifested in the design of the counter, showed that such cells could be used to make a registration plane for drift time-projection chamber.

\section{Detector construction}
\label{sec:Detector}

A schematic view of the test matrix counter (TMC) cross-section is shown in
Fig.1.
The design of the counter is assembled on three parallel mounting racks, passed through holes in a set of flat copper rings (6 pieces) with an outer diameter of 73 mm and a width of 8 mm. The rings are evenly distributed along the length of the racks with a pitch of 10 mm and are isolated from each other and from the metal of the racks with fluoroplastic insulators. The upper end of the mounting racks in the figure is fixed on a solid disk, which is a high-voltage electrode for creating a drift field. The lower end of the racks is fixed on a solid mounting disk with an installed cage of matrix counters.
\begin{figure}
    \hspace{5.60pc}
	\hspace{0cm}\includegraphics[width=1.0\columnwidth,angle=0.]{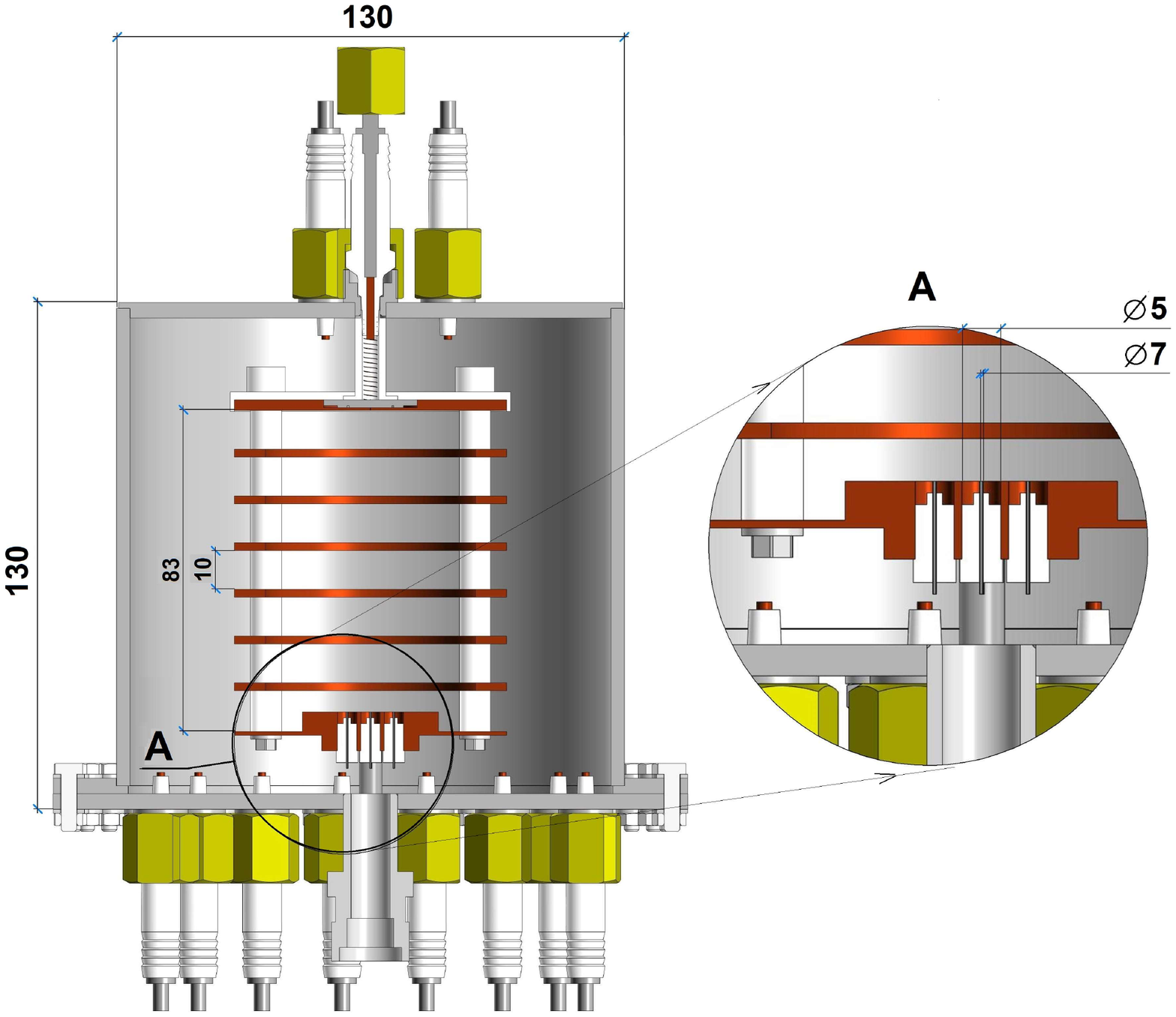}
	\caption{Schematic view of the test matrix counter. \textbf{A}: the $3\times3$ matrix.}
	\label{fig:detector}
\end{figure}

A set of butt-end counters placed in holes (d=5 mm) passed through
the holder and looked as $3\times3$ matrix. A thickness of a wall between
adjacent cells is 1.0 mm. The cells are numbered as (1,2,3) - first row,
(4,5,6) - second row and (7,8,9) - third row. The axis of the central cell
(No.5 ) coincides with the axis of the detector.
A cylindrical Teflon insulator with a stainless steel tube of 0.4 mm inner and 0.7 mm outer diameters installed along the axis is inserted into the holder hole to make a separate cell.
A working butt end of the tube is grinded and electropolished before set into the insulator.
This end flush with the holder plane and is the cell's anode.
The backside of the tube is
connected with a central electrode of a spark plug which is installed
hermetically through a set-up body flange. Other set-up features are shown in
Fig.1.

\section{Detector's working characteristics}
\label{subsec:characteristics}

A source of the $^{238}$Pu with $3.8\times10^4$ Bq activity was used for the
detector's amplitude characteristics investigation. Energies of the source
$\alpha$-particles are equal to 5456 keV (28.3\%) and 5499 keV (71.6\%). The
isotope is a source of the conversion electrons with energies of 22.53 keV
(10.80\%), 26.31 keV (9.48\%), 39.18 keV (5.68\%), 43.48 keV(2.15\%) and
characteristics  X-ray quanta with energies of 13.60 keV (3.97\%), 17.06 keV
(5.57\%) and 20.29 keV(1.28\%) \cite{nudat3,xdb}.

Measurements were made primarily with the $\alpha$-particles energy releases to simplify a search and identification of energy lines.
For this, the open-source surface was used, visible through the window ($d = 0.8$ mm) made in the center of the bottom of the cylindrical recess ($h = 0.3$ mm) of the high-voltage electrode, designed to accommodate the source.
An $\alpha$-particle flux through this window corresponds to the
$\sim$240 Bq source activity.  The axis of the source and the central cell
coincide. Measurements were made for the (\emph{I}) 96.3\% Ar + 3.7\% Xe and (\emph{II})
90\% Ar + 10\% CH$_4$ gas mixtures at the 620 and 62 Torr pressures. A pressure
620 Torr is normal for the BNO INR RAS location height. Pulses were taken by a
charge sensitive preamplifier (CSA)  and recorded by a digital oscilloscope
(DO).
\begin{figure}[pth]
\centering
\vspace{-7.0pc}
\includegraphics*[width=0.55\textwidth]{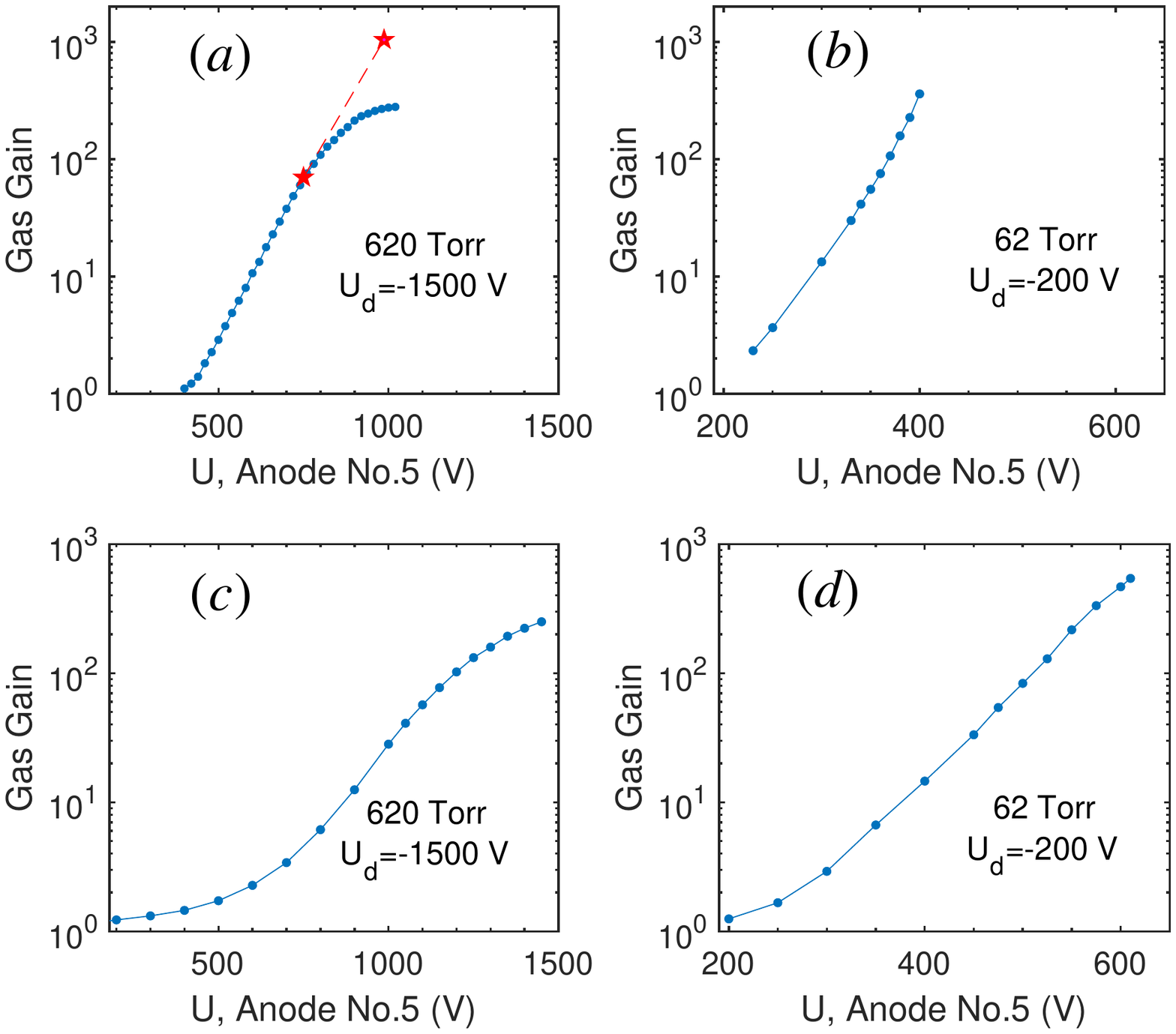} \vspace{-6.0pc}
\caption{Dependence of the gas multiplication factor on the voltage on cell No.5, filled with various gas mixtures. For the gas mixture \emph{I} (96.3\% Ar + 3.7\% Xe) at a pressure and drift voltage:
(\emph {a}) - [P = 620 Torr, V$_d$=-1500 V]; (\emph{b}) - [P=62 torr, V$_d$=-200 V];
star symbols are GMF values calculated from comparing the energies and amplitudes of impulses from photons and $\alpha$-particles.
For a gas mixture \emph{II} (90\% Ar + 10\% CH$_4$) at a pressure and drift voltage: (\emph{c}) - [P=620 Torr, V$_d$=-1500 V ]; (\emph{d}) - [P=62 torr, V$_d$=-200 V].}
\label{fig:gas_gaine}
\end{figure}

Figure 2 shows the voltage dependence of the gas gain factor (GMF) of cell No. 5(central) for the gas mixture (\emph{I}) at two different pressures.
The dependence of a GMF on a high voltage value (V$_a$) at a pressures of 620 Torr is shown in Fig.2(\emph{a}). A voltage (-1500 V) is fided on the drift electrode (Vd).
The anodes of the remaining cells are grounded. The kink in the characteristic at voltages above $\sim$800 V is
associated with an output voltage limitation of the used CSA.

Figure 3(\emph{a}) shows
the amplitude spectra from cell No.5 at operating voltage U$_a$=700 V and
drift voltage (-1500 V): spectrum \emph{1} - surrounding anodes are grounded;
spectrum \emph{2} - at the surrounding anodes 700 V. Spectrum \emph{3} was obtained with the
parallel combination of all cells into one counter.
\begin{figure}
\centering
\vspace{-12.0pc}
\includegraphics[width=0.65\textwidth]{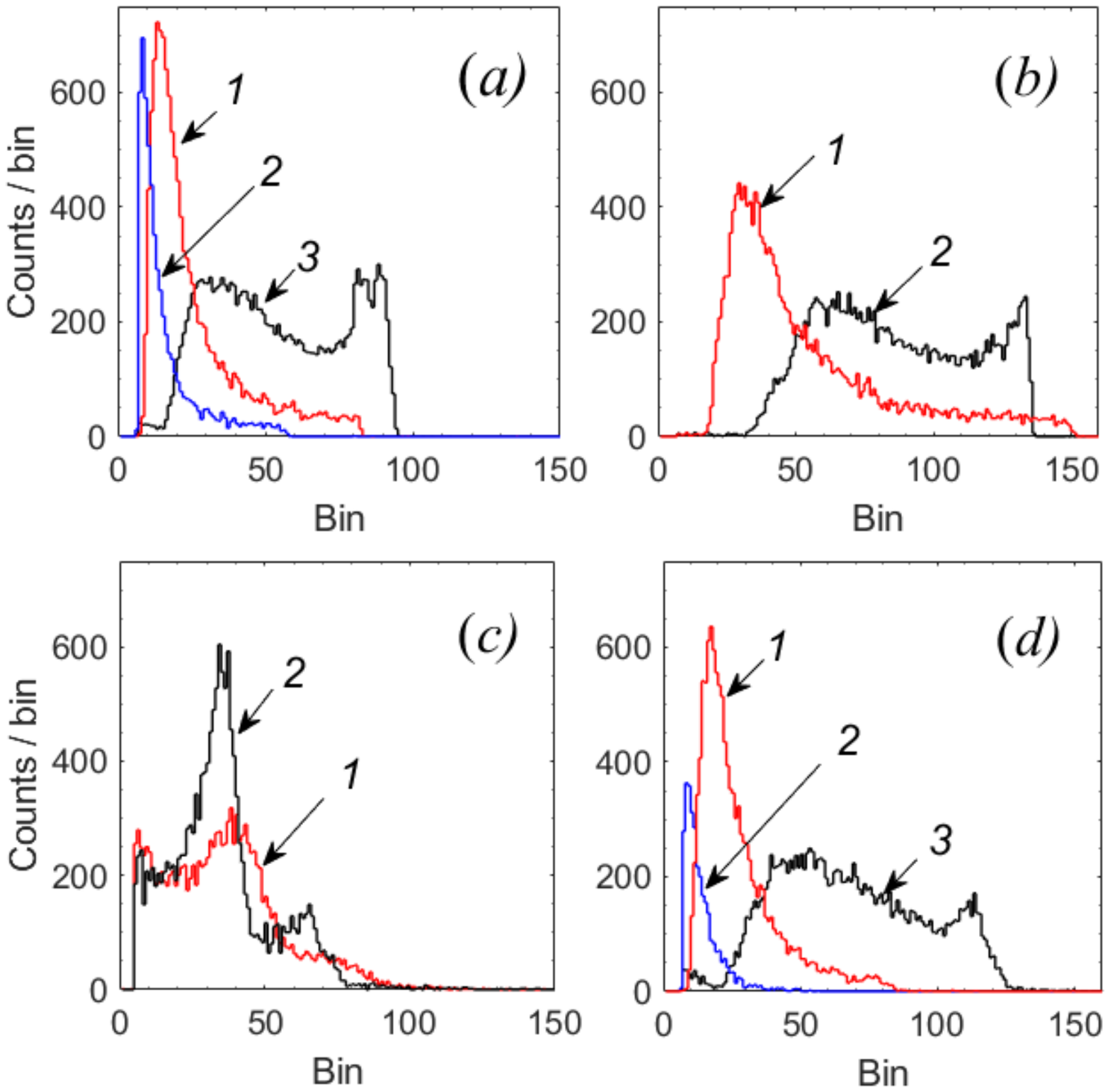}\vspace{-7.0pc}
\caption{
Pulse-height spectra at dependence from the anode and drift voltage for cell No.5 filled with the gas mixture (I), 96.3\% Ar + 3.7\% Xe.
(\emph{a}) - at V$_a$=700 V, V$_d$=-1500 V, and P=620 Torr: spectrum \emph{1} - all surrounding anodes are grounded;
spectrum \emph{2} - voltage 700 V fed at all anodes;
spectrum \emph{3} - all cells integrated into the single counter by parallel  connection.
(\emph{b}) - at V$_a$=750 V and P=620 Torr, all surrounding anodes are grounded:
spectrum \emph{1} at V$_d$=-1500 V;
spectrum \emph{2} at V$_d$=-200 V.
(\emph{c}) - with  $^{238}$Pu covered by Ni-foil 50 $\mu$m thickness, V$_a$=987 V, all surrounding anodes are grounded:
spectrum \emph{1} at V$_d$=-1500 V;
spectrum \emph{2} at V$_d$=-200 V.
(\emph{d}) -  at V$_a$=370 V and V$_d$=-200 V, P=62 Torr:
spectrum \emph{1} - all surrounding anodes are grounded;
spectrum \emph{2} - voltage 370 V fed at all anodes;
spectrum \emph{3} - all cells integrated into the single counter by parallel connection.
}
\label{fig:spc3}
\end{figure}

The range of an $\alpha$-particle with the 5.5 MeV in the gas mixture (\emph{I})  at
620 Torr is equal to 5.7 cm according to calculations made by using the data
from \cite{Nemets75}. The range goes entirely in the counter working volume. A spectrum
shape will reflect a distribution of energy release projections to the
registration plane transformed accordant to a value of the cell gas gain and
its uniformity around the anode circle. A peak with a double top is presented
in spectrum \emph{3} at the maximal energy release. A view of the peak could
reflect the distribution of the GMF in the cells around the central one. All
perimeter cells will have a GMF larger than the central one.
Moreover, the corner cells will have a GMF larger than the side ones. Since
the structure under study does not have a screening grid, the drift
electrode's potential should affect the cells' GMF. Energy release spectra of
the cell No.5  were measured at V$_a$=750 V for two values of V$_d$ to estimate this
influence. All other cells were grounded. Measured results are shown in Fig.3(\emph{b}).
Spectrum \emph{1} was measured for V$_d$=-1500 V, and spectrum \emph{2} was measured for V$_d$=-200 V.
It is seen that increasing of the V$_d$ at 1300 V decreases GMF at $\sim$11\%. A
peak at the maximal energy releases is seen in spectrum \emph{2}. Its appearance
could be explained by increasing the effective area of primary ionisation
electrons collection to the anode under the action of the own electric
potential due to decreasing drift electric field and increasing electron drift
time.
The number of $\alpha$-events with a total energy release will increase too.

The TMC response on low energy radiation was obtained by using the $^{238}$Pu
source covered with the 50 $\mu$m Ni-foil. The residual source X-rays and
nickel characteristic photons (7.5 keV) excited by the source
$\alpha$-particles, electrons and X-rays are coming from the foil surface. The
spectra obtained at V$_a$=987 V and V$_d$=-1500 V (spectrum \emph{1}), V$_d$=-200 V (spectrum
\emph{2}) are shown in Fig.3(\emph{c}).
The 7.5 keV ($\sim$36 bin) and $\sim$17 keV (66 bin) lines are visible
on the spectra. An energy resolution of the 7.5 keV line on spectrum \emph{1} is
$\sim$26\%.

The resolution is worsening with the drift field growth because
the working point of the cell drifted deeper to the limited proportionality
region. Such a conclusion could be made because a 7.5 keV peak appears in the $\sim$103 bin at V$_a$= 1025 V and V$_d$=-200 V, but
17 keV is in 138 bin only. The peaks have flowed together at V$_d$=-1500  V,
and their common top is at $\sim$134 bin. A 7.5 keV line (A$_{7.5}$) value
was used to plot the GMF dependence in the region of extremely acceptable
working voltages.
The GMF values were calculated from comparing energies and amplitudes of the
$\alpha$-particles and photons pulses by formula
GMF$_{7.5}=A_{7.5}/A_{5499})\times(5499/7.5)\times$ GMF$_{5499}$. Obtained
values 
are plotted in Fig.2(\emph{a}) by star points. The two parts of the dependence are connected with a straight dashed line.

The matrix counter potentially could be used in the dark matter search experiments where energy and spatial characteristics of the nuclear recoil tracks are recorded simultaneously as in the experiment with using a low-pressure gaseous time projection chamber (NEWAGE-0.3b) \cite{Yakabe2020}.
In this case, the measurements should be carried out at low pressure to obtain a sufficiently extended trajectory.

To obtain information on the parameters of the MMPC performance, we carried out test measurements at a gas pressure of 62 Torr.
The measurements were made at the 62 Torr gas pressure to get information
about working characteristics possible parameters.
The GMF dependence on the high voltage for the cell No.5  filled up to 62 Torr with the mixture (\emph{I}) is shown in Fig.2(\emph{b}). A potential of the drift electrode is
V$_d$=-200 V. All other anodes were grounded to get this curve. Figure 3(\emph{d}) shows
the amplitude spectra from cell No. 5  at anode voltage U$_a$=370 V and a drift
voltage U$_d$=-200 V: spectrum \emph{1} was measured when all surrounding anodes
were grounded; spectrum \emph{2} was measured when at the surrounding anodes 370 V.
Spectrum \emph{3} was obtained with the parallel combination of all cells into one
counter.
The 5.5 MeV $\alpha$-particle range in the counter working gas is 57 cm. An
energy release at a distance of 7.8 cm between the drift electrodes and the
registration plane is 752 keV if the energy losses assume uniform along a
track. The peak at the end of spectrum \emph{3} corresponds to this energy release.

The gas mixture (\emph{I}) regards the non-metastable Penning mixture \cite{Kubota70}. A reduced
working voltage is observed for this mixture compared to the usual
(Ar+CH$_4$)-mixture to reach the identical GMF.
Reducing the operating voltage can be helpful to increase the detector's reliability with small gaps between the working electrodes.
Measurements with the gas mixture (\emph{II}) were made to specify a value of an
expected gain. The 5.5 MeV $\alpha$-particles ranges are similar in both
mixtures.

\begin{figure}
\centering
\vspace{-15.0pc}
\includegraphics*[width=0.65\textwidth]{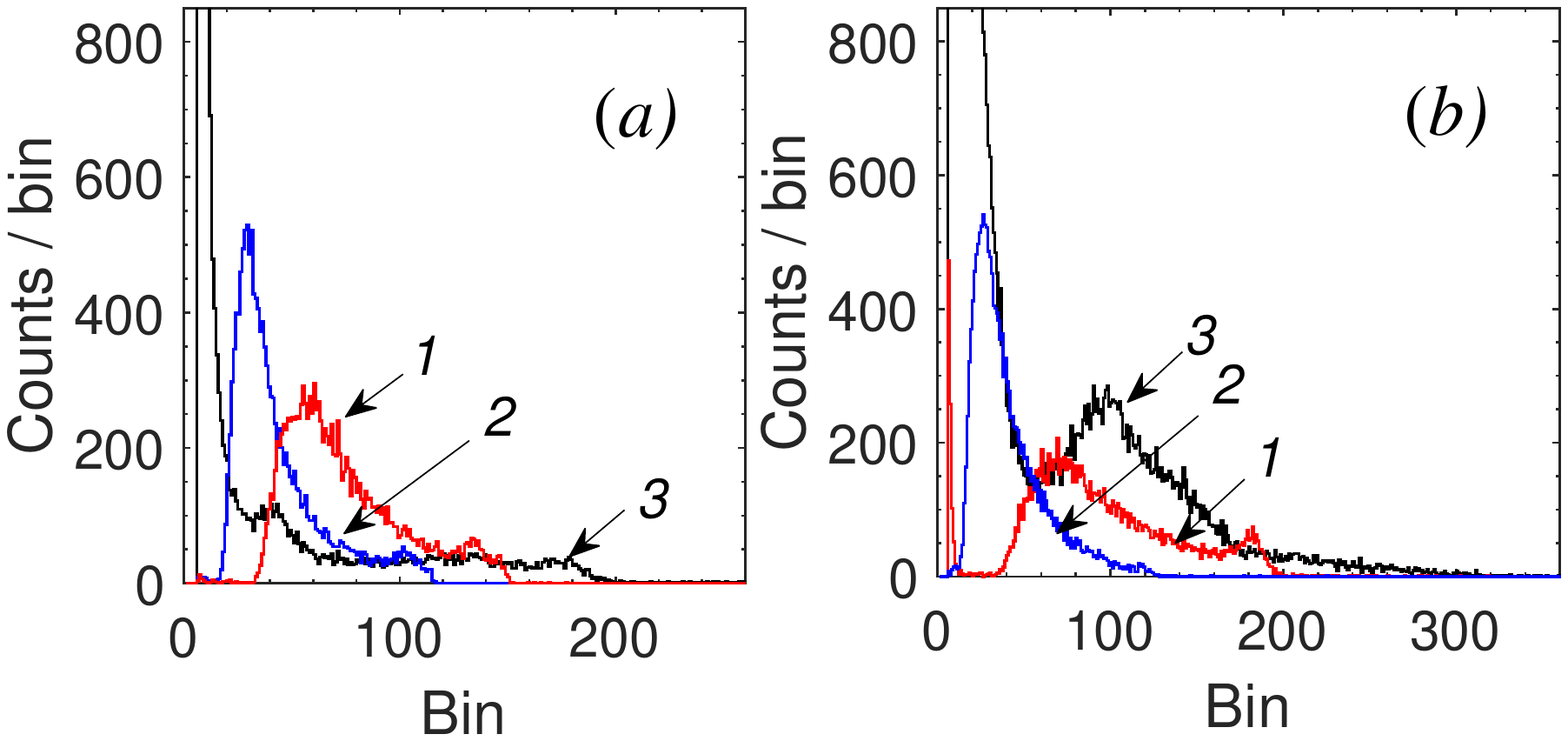}\vspace{-13.0pc}
\caption{
Pulse pulse-height spectra from cell No.5. filled with the gas mixture (\emph{II}), 90\% Ar + 10\% CH$_4$.
(\emph{a})- at V$_a$=1250 V and V$_d$=-1500 V for P=620 Torr: spectrum \emph{1} - all surrounding anodes are grounded;
spectrum \emph{2} - voltage 1250 V fed at all anodes;
spectrum \emph{3} - all cells integrated into the single counter by parallel connection.
(\emph{b}) - at V$_a$=575 V and V$_d$=-200 V for P=62 Torr:
spectrum \emph{1} - all surrounding anodes are grounded;
spectrum \emph{2} - voltage 575 V fed at all anodes;
spectrum \emph{3} - all cells integrated into the single counter by parallel connection.
}
\label{fig:spc4}
\end{figure}
The GMF dependence on the high voltage for the cell No.5  filled up to 620
Torr with the mixture (\emph{II}) is shown in Fig.2(\emph{c}). A potential of the drift
electrode is V$_d$ $=-1500$ V. All other anodes were grounded to get this curve.
Amplitude spectra from the cell No.5  for the V$_a$=1250 V and V$_d$=-1500 V  are
shown in Fig.4(\emph{a}) where spectrum \emph{1} was measured when all surrounding anodes were
grounded, the spectrum \emph{2} was measured when the voltage 1250 V was fed at all
anodes. Spectrum \emph{3} was obtained when all cells were integrated into the single
counter by a parallel connection.

It is seen from a comparison of the GMF dependences for the gas mixtures (\emph{I})
and (\emph{II}) at 620 Torr that GMF=100 value obtained at V$_a$=790 V and V$_a$=1200 V
correspondingly. The same GMF value for the 62 Torr was obtained at V$_a$=370 V
and V$_a$=510 V. A voltage gain for the gas mixture (\emph{I}) is significant. It is
seen from a comparison of the GMF dependences for the gas mixtures (\emph{I}) and (\emph{II})
at 620 Torr that GMF=100 value obtained at V$_a$=790V and V$_a$=1200 V
correspondingly. The same GMF value for the 62 Torr was obtained at V$_a$=370 V
and V$_a$=510 V. A voltage gain for the gas mixture (\emph{I}) is significant.

It is seen from Fig.3(\emph{a}) that a supply of the V$_a$=700 V on the before grounded
cells around the cell No.5  decreases an amplitude at $\sim$29\%. It
corresponds to decreasing of a voltage at the cell No.5 up to $\sim$675V.

A GMF on the peripheral cells should be larger than on the cell No.5  because
of the absence of other anodes with high voltage potential outside the matrix.
This fact should be taken into account as an explanation of the behavior of
the $\alpha$-peak shape and position. One can expect that the GMF of a cell in
a larger dimension matrix should decrease from the maximum up to any constant
value when a cell position shifts from a periphery to the center. A
distribution of GMF values among cells could be more complicated in a region
of a drift field forming electrodes potentials influence.

\section{Discussion of results}

The field strength determines the gas multiplication coefficient of an individual cell of the TMC at the outer edge of the annular end of the anode tube. This tension is a superposition of the tensions created by the potentials of all metal parts surrounding the selected cell. In this case, the negative potentials of the drift electrode, forming rings, and grounded parts increase the strength of the electric field, while the positive potentials of the anodes of the surrounding cells decrease it. The magnitude of the tension can vary along the circumference depending on the total tension created by the unevenly distributed surrounding potentials. This nonuniformity has little effect on the averaged GMF of the cell at the used working gas pressure, judging by the resolution at 7.5 keV.
Preliminary measured or calculated correction could be made to decrease the influence of separate cells GMF values differences to an energy resolution of an integrated spectrum.

Using a tube anode instead of a solid cylindrical end gives a voltage gain to achieve the same values of the GMF.  A modelling calculation shows that the maximal value of field strength at the outside edge of an anode tube butt-end will be at $\sim$12\% larger than the one of a bar anode. This difference gives the sought for voltage gain to obtain similar GMF values for the tube anode and the bar one.

Additional measurements of GMF were made with an anode butt-end dipped at
0.5 mm below (position 1) the operational level (position 2) of the anode in
the copper platform surface plane. The one lifted at 0.5 mm above (position 3)
the level (position 2) to estimate an influence of the butt-end installation
accuracy along a cell axis on the counter characteristics. The measurements
were carried out on mixture (\emph{I}) at a pressure of 620 Torr, V$_a$=800 and V$_d$=$-1800$ V.
In case (1), the GMF decreased to a value of 0.632
of the GMF value for option (2), which is equivalent to a decrease in the
operating voltage at point (position 2) to 754 V or by 5.7\%.
In case (3), the GMF also decreased to 0.936, equivalent to a decrease in the
voltage at point (position 2) by 1.1\%.

The variation coefficients $(k)$ for the examined particular case will be equal to $kd=0.736$ mm$^{-1}$ for a dipping and $ke=0.128$ mm$^{-1}$ for an extending if one assumes that a GMF value varies arcwize with the specified position deviations. The adduced data allow one to estimate an influence of a possible deviation of the butt-end plane from the normal one on GMF value distribution around the anode ring. One point of the anode ring will be in the platform surface plane (position 2), but the diametrically opposite point will be in the (position 1) at the ``$h$'' depth if the anode plane is not a normal tube axis. A relative GMF value $(M)$ will vary from $M_2=1$ in the (position 2) up to $M_1=M_2\times(1-h \times kd)=0.963$ in the (position 1) if one take h=0.05 mm. GMF value deviation could impair the energy resolution of the detected energy releases.
One should keep the requirement of an anode butt-end plane perpendicularity to the tube axis during the anode preparation to exclude this factor.

\section{Conclusion}
\label{Conclusion}
The results presented in this paper show that a cell consisting of a 0.7 mm medical needle tube inserted into the centreline of a Teflon dielectric moulded in a 5 mm cylinder tube on a copper plate may be used as a gas proportional counter. The tube ring butt-end is an anode. An energy resolution of $\sim$26\% at 7.5 keV was obtained for the 96.3\% Ar + 3.7\% Xe gas mixture at the 620 Torr pressure. A multicell counter in the form of an arbitrary size matrix could be made on a base of such cells.
It is shown on the $3\times3$ matrix sample that a gas multiplication coefficient of the central cell depends on electric field potentials presented at peripheral cells, drift and field-shaping electrodes, and surrounding metallic parts of the counter construction. The GMF values differ from the constant at different matrix sections due to this reason.
This difference could be eliminated in the process of a summation of signals from different cells using the compensating coefficients. A condition of an anode butt-end plane perpendicularity to the tube axis during the anode preparation should keep preventing an influence of an anode butt-end geometry to the counter energy resolution. The simplicity of separate cell preparation and absence of any technical limitations of the counter registration surface shape allows one, in principle, to construct a multicell sensitive surface in the form of the sphere. Such construction could represent an interest as a central electrode of the spherical proportional chamber \cite{Katsioulas20} since it allows to obtain a space angular sensitivity in the registered signals analysis.

The work is carried out in accordance with the INR RAS research plan.

\bibliographystyle{ieeetr}


\end{document}